\documentclass{article}
\usepackage{amsmath}
\usepackage{footmisc}
\usepackage{graphicx}
\usepackage{amssymb}
\usepackage{enumerate}
\usepackage{multirow}
\usepackage{longtable}
\usepackage{booktabs}
\usepackage{multicol}
\usepackage{amsthm}
\usepackage{graphicx}
\begin{document}

\cleardoublepage \pagestyle{myheadings}
\bibliographystyle{plain}
\title{Alice-Bob Physics: Coherent Solutions of \\
Nonlocal KdV Systems}
\author{S. Y. Lou$^{1,3}$\thanks{Email: lousenyue@nbu.edu.cn} 
and Fei Huang$^{2,3}$ 
\\
\footnotesize $^{1}$ \it Shanghai Key Laboratory of Trustworthy Computing, East China Normal University, Shanghai 200062, China,\\
\footnotesize $^{2}$\it
Physical Oceanography Laboratory, Department of Marine
Meteorology, Ocean University of China, Qingdao, 266100, China
\\
\footnotesize $^{3}$\it Ningbo Collabrative Innovation Center of Nonlinear Harzard System of Ocean and Atmosphere\\
\footnotesize \it and Faculty of Science, Ningbo University,  Ningbo, 315211, China}
\maketitle

\begin{abstract}
In natural and social science,  many events happened at different space-times may be closely correlated. Two events, $A$ (Alice) and $B$ (Bob) are defined correlated if one event is determined by another, say, $B=\hat{f}A$ for suitable $\hat{f}$ operators. Taking KdV and coupled KdV systems as examples, we can find some types of models (AB-KdV systems) to exhibit the existence on the correlated solutions linked with two events. 
The idea of this report is valid not only for physical problems related to KdV systems but also for  problems described by arbitrary continuous or discrete models. 
The parity and time reversal symmetries are extended to shifted parity and delayed time reversal. The new symmetries are found to be useful not only to establish  AB-systems but also to find group invariant solutions of numerous AB-systems. A new elegant form of the $N$-soliton solutions of the KdV equation and then the AB-KdV systems is obtained. 
A concrete AB-KdV system derived from the nonlinear inviscid dissipative and barotropic vorticity equation in a $\beta$-plane channel is applied to the two correlated monople blocking events which is responsible for the snow disaster in the winter of 2007/2008 happened in Southern China.
 \\
\bf PACS numbers: 
\rm  
05.45.Yv 
02.30.Ik 
11.30.Er 
11.10.Lm	
03.50.Kk 
02.30.Jr
02.30.Ks
\end{abstract} 

\leftline{\bf Additional Information}
\leftline{\bf Competing financial interests: \rm The authors declare no competing financial interests.}

\section{Introduction}
It is known that various events happened in different places and times are closely correlated and/or entangled\cite{ABW}. 
For example, the decrease of ice at Arctic sea in the summer of 2007 (event $A$, $A=A(x,t)$) is considered to be responsible for the heavy freezing rain in winter 2008 in Southern China (event $B$, $B=B(x',t')=\hat{f}A$)\cite{08}. The strong El Nino event occurred in 1997 (event $A$) induced the heavy Yangtze River flooding in 1998 (event $B(x',t')=\hat{f}A$)\cite{98}. In the communication field, in order to safely get information from Alice (event $A$), Bob should have a private Key $\hat{f}$ such that he can find the thing he wants ($B=\hat{f}A$). It is reported that the event $B$, the detection of a signal of gravitational wave, is due to the event $A$, the merging of two black holes from 1.3 billion light years away\cite{GW}; 
In quantum physics, many faraway particles (atoms) may construct a completely entangled state, a measurement of one particle (event $A$) will affect the state of the other (event $B=\hat{f}A$)\cite{ENT}. 

Traditionally,  physical models are locally established around a single space-time point, say, $\{x,\ t\}$. To describe the correlated events happened in two or more places, one has to establish some kinds of new physically significant models. In this paper, we restrict ourselves to study only two-place physical models, Alice-Bob (AB) models for simplicity. We say Alice-Bob physics if the physics is related to 
two correlated/entangled events occurred in two places $\{x,\ t\}$ and 
$\{x',\ t'\}$ respectively. The event at 
$\{x,\ t\}$ is called Alice event (AE) (denoted by $A(x,\ t)$) 
and the event at $\{x',\ t'\}$ is called Bob event (BE) (denoted by $B(x',\ t')$).  
The events AE and BE are called correlated/entangled if AE happens, the BE can be determined at once by the correlation condition
\begin{equation}
B(x',\ t')=f(A)=A^f=\hat{f}A,
\end{equation}
for suitable operator(s) $\hat{f}$. 

Usually, $\{x',\ t'\}$ is not neighbour to $\{x,\ t\}$. Thus, the intrinsic two-place models, the Alice-Bob systems (ABSs), are nonlocal. Some special types of two-place nonlocal models have been proposed, say, the nonlocal nonlinear Schr\"odinger (NLS) equation (called also PT symmetric NLS equation), 
$$
iA_t+A_{xx}\pm A^2B=0,\  B=\hat{f}A=\hat{P}\hat{C}A=A^*(-x,t), $$
was firstly proposed by  Ablowitz and Musslimani\cite{AM}. The operators     
$\hat{P}$ and $\hat{C}$ are the usual parity and charge conjugation. Some types of other nonlocal nonlinear systems such as the coupled nonlocal NLS systems \cite{DNLS}, the nonlocal modified KdV systems \cite{MKdV}, the discrete nonlocal NLS system \cite{dNLS} and the nonlocal Davey-Stewartson systems\cite{DS} were also proposed. Though the nonlocal NLS and the other known models mentioned above have been studied by many authors, they are all proposed in the mathematical way from the integrability requirement without any possible physical applications. 

In this paper, we attempt to provide a possible physical application: They are special AB systems and can be possibly used to describe some special types of AB physics. In section 2, we discuss the general aspect of the AB systems via shifted parity and the delayed time reversal. In section 3, the detailed group invariant solutions for general AB-KdV systems and symmetry breaking solutions for a special AB-KdV system are studied. 
The multiple soliton solutions of the usual integrable systems have been applied in almost all the physical fields and there are some types of formally different complicated expressions such as the Hirota forms, determinant forms, Pffafion forms and so on \cite{Nsoliton}. Here, a much more elegant form of the solutions for the KdV type systems is presented. 
In section 4, a special AB-KdV system is derived from a rotating fluid system via multiple scale method. A symmetry breaking soliton solution of the special KdV system is applied to a two-atmospheric-blocking-events which is related to the heavy snow disaster  in the winter of 2007/2008 in the south of China. The last section is a summary and discussion. 

\section{General aspect of AB systems}

There exist two possible ways to establish ABSs which may be used to describe AB physics in the future. 

 (i). The first way is to solve out two events AE and BE from traditional basic principle equation(s),
\begin{equation}
F(u)=0\ \label{Fu}
\end{equation}
by assuming 
\begin{equation}
u=U(A(x,t),\ B(x',\ t')) \label{Uu}
\end{equation}
for suitable $U$, where $u$ may be a scalar, a vector,\ a matrix or a tensor. 
Eq. \eqref{Fu} may be arbitrarily known principle systems, say, the Navier-Stokes (NS) equation for AB fluid problems, the Maxwell system for electro-meganetic-optic AB problems, the Schr\"odinger equation in quantum physics, the Einstein's equation for the gravitational AB problems and the various types of their generating models such as the Korteweg de-Vrise (KdV) equation, the NLS equation, the Kadomtsev-Petviashvilli (KP) equation and so on. If the original principle system \eqref{Fu} is classical, then the derived AB systems should also be classical. Meanwhile, if \eqref{Fu} is a quantum system, say the Schr\"dinger equation or Dirac equation, the derived AB systems are naturally quantum. 

(ii). The second way is to establish or derive new models by introducing suitable  correlated/entangled conditions for known coupled systems, for instance, the nonlocal NLS system obtained by Ablowitz and Musslimani
\cite{AM} is just derived from a coupled system, the Ablowitz-Kaup-Newell-Segue (AKNS) system. More examples of AB systems derived from coupled equations have been offered in the preprint paper \cite{ABS}.

To find out some concrete results, we 
introduce a principle \cite{ABS}, the AB-BA equivalence principle, which requires $U(A,\ B)=U(B,\ A)$ and 
\begin{eqnarray}
&& B=\hat{f}A,\quad A=\hat{f}B,\quad \hat{f}^2=1.\label{ABBA}
\end{eqnarray}
Substituting the two event  assumption \eqref{Uu} into the original principle \eqref{Fu} and splitting the resulting equation  to two, we may have
\begin{eqnarray}
&&F_1(A,\ B)=0,\quad F_2(A,\ B)=0.\label{F12}
\end{eqnarray}
Two equations of \eqref{F12} should be  reduced back to  one equation because of the correlated condition \eqref{ABBA}. To reduce two equations of \eqref{F12} to one we can apply the operator $\hat{f}$ on one of them. The results should be 
\begin{eqnarray}
&&\hat{f}F_2(A,\ B)=F_1(A,\ B),\ F_2(A,\ B)=\hat{f} F_1(A,\ B)\label{F1F2}
\end{eqnarray}
with the condition \eqref{ABBA}.

Generally, there are infinitely many solutions of $\hat{f}^2=1$. For instance, any solution of $F(x,\ x')\pm F(x'\ x)=0$ for arbitrary two-dimensional function  $F$ solves $\hat{f}^2=1$ if we express 
$x'=\hat{f}x$. For linear transformations of $N$ variables 
$x=(x_1,\ x_2,\ \ldots,\ x_N)^T$, the superscription $T$ of a matrix means the transposition of the matrix, we have the following theorem and conjecture. \\
\bf Theorem. \rm \em 
The 
linear transformation 
\begin{equation}
x'=M x+x_0= \hat{f}x, \ x=M x'+x_0= \hat{f}x', \label{ABL} 
\end{equation} 
possesses the solution 
\begin{equation}
x'_i=\left(\begin{array}{cc}
\Gamma\Omega-I_n & \Gamma\\
2\Omega-\Omega\Gamma\Omega & I_{N-n}-\Omega\Gamma
\end{array}\right) x_i+\left(\begin{array}{c}
x_{0n} \\ -\Omega x_{0n}
\end{array}\right),\ n=0,\ 1,\ \ldots,\ N, \label{SABL} 
\end{equation} 
where $x_i=(x_{i_1},\ x_{i_2},\ \ldots,\ x_{i_N})^T$ is related to all the possible permutations of $\{i_1,\ i_2,\ \ldots,\ i_N\}=\{1,\ 2,\ \ldots,\ N\}$, $\Gamma,\ \Omega$ and $x_{0n}$ are arbitrary $n\times (N-n)$,\ $(N-n)\times n$ and $n\times 1$ matrices, respectively, $I_n$ and $I_{N-n}$ are $n\times n$ and $(N-n)\times (N-n)$ unit matrices. 
\rm 
\\
\bf Conjecture.  \rm  All solutions of \eqref{ABL} have the form \eqref{SABL}.
\\
\bf Remark. \rm The theorem can be simply proved by checking the condition $M^2=I$ and $Mx_0=-x_0$. The conjecture has been directly checked for $N\leq 4$ in \cite{ABS}. We believe the conjecture is also true for $N>4$.
(ii). For $n=0$,\ $\Gamma=\Omega=0$ the transformation \eqref{SABL} is a trivial identity transformation. 
For $n=N$,\ $\Gamma=\Omega=0$ the transformation \eqref{SABL} is a shifted parity transformation for all variables $x_i,\ i=1,\ 2,\ \ldots,\ N$. (iii). Because any similarity transformation $\{x',\ x,\ x_0, M\}\rightarrow \{Ux',\ Ux,\ Ux_0,\ UMU^{-1}\}$ is also a solution of \eqref{ABL}, the conjecture implies that the solution \eqref{SABL} is similarity transformation invariant up to redefinition of the arbitrary matrices $\Gamma,\ \Omega$ and $x_{0n}$. 

\section{AB-KdV systems and their coherent structures}    
To make it more clear, we take the well known KdV equation \cite{KdVa}
\begin{eqnarray}
&&KdV\equiv u_t+u_{xxx}+6uu_x=0
\label{kdv}
\end{eqnarray}
as a simple example. 

The KdV equation was first introduced by Boussinesq \cite{Bq} and rediscovered by Diederik Korteweg and Gustav de Vries \cite{KDV}. The KdV equation has various connections to physical problems. It approximately describes the evolution of long, one-dimensional waves in many physical settings, including
shallow-water waves with weakly non-linear restoring forces,
long internal waves in a density-stratified ocean,
ion acoustic waves in a plasma,
acoustic waves on a crystal lattice\cite{KdVa} and the 2-dimensional quantum gravity\cite{QG}.
The KdV equation can be solved using the inverse scattering transform and other methods such as those applied to other integrable systems \cite{IST}.

Furthermore, for simplicity, we take the symmetric $U$ as a simplest form $U=\frac12(A+B)$. 
From the equation
\eqref{kdv}, it is clear that the KdV equation is parity-time reversal invariant, space-time translation invariant and complex conjugate invariant. Thus, we can take $\hat{f}$ as 
\begin{equation}
\hat{f}=\hat{P}_s\hat{T}_d,\ \hat{f}A=A(-x+x_0,\ -t+t_0)=B, \label{PT}
\end{equation}
for the real KdV system and $\hat{f}=\hat{P}_s\hat{T}_d\hat{C},\ \hat{f}A=A^*(-x+x_0,\ -t+t_0)=B$
for the complex KdV equation, where 
$\hat{P}_s$ 
is a shifted parity defined by 
$\hat{P}_sx=-x+x_0$, $\hat{T}_d$ 
is a delayed time 
reversal defined by $\hat{T}_dt=-t+t_0$, and $\hat{C}$ is the usual charge conjugate. 

Substituting $U=\frac12(A+B)$ into the KdV equation \eqref{kdv}, we have
$$A_t+B_t+A_{xxx}+B_{xxx}+3(A+B)(A_x+B_x)=0$$
which can be split to two equations
\begin{eqnarray}
&&A_t+A_{xxx}+3(A+B)A_x+G(A,\ B)=0,\label{kdvA}\\
&&B_t+B_{xxx}+3(A+B)B_x-G(A,\ B)=0,\label{kdvB}
\end{eqnarray}
where $G(A,\ B)$ may be an arbitrary functional of $A$ and $B$. Applying $\hat{f}$ defined in Eq.  \eqref{PT} on Eq. \eqref{kdvB} will lead to a compatibility condition 
\begin{eqnarray}
G(A,\ B)=\hat{f}G(A,\ B).\label{GG}
\end{eqnarray}
In other words, $G(A,\ B)$ is not an arbitrary functional but an arbitrary $\hat{f}$ invariant functional.

Finally, we obtain a quite general nonlocal AB-KdV system 
\begin{eqnarray}
ABKdV\equiv A_t+A_{xxx}+3(A+B)A_x+G(A,\ B)=0, \qquad B=A(-x+x_0,\ -t+t_0)\label{19}
\end{eqnarray}
or $B=A^*(-x+x_0,\ -t+t_0)$ with an arbitrary $\hat{f}$ invariant  functional $G(A,\ B)$. 

It is known that 
the original KdV  equation is a famous integrable model. However, the integrability of the AB-KdV system \eqref{19} is not very clear because the introduction of the arbitrary $\hat{f}$ invariant functional $G(A,\ B)$ though there are some different formal Lax pairs. The first formal Lax pair possesses the form,
\begin{eqnarray}
&&\psi_{xx}=-\frac12\left(\begin{array}{cc}
A-\lambda & B-\lambda \\ B-\lambda & A-\lambda
\end{array}\right)\psi,\label{pxx}\\
&& \psi_{t}=\left(\begin{array}{cc}
\frac12(A-h)_x+(h-A-2\lambda)\partial_x & \frac12(B+h)_x-(h+B+2\lambda)\partial_x \\ \frac12(B+h)_x-(h+B+2\lambda)\partial_x & \frac12(A-h)_x+(h-A-2\lambda)\partial_x
\end{array}\right)\psi, \label{pt}
\end{eqnarray}
where 
\begin{equation}
h=\Phi^{-1}[G(A,B)+3B(B-A)_x]_x, \ \Phi=(A-B)_x\partial_x^{-1}+2(A-B)+\partial_x^2,
\end{equation}
and $\Phi$ is just the recursive operator of the classical KdV equation with the potential $A-B$.

If the arbitrary $\hat{f}$ invariant functional $G$ is fixed, one may obtain more concrete Lax pairs, say, if we select
\begin{equation}
G=0,\ (A\pm B)(B\mp A)_x,\label{25}
\end{equation}
 the related AB-KdV systems are just $ABKdV_i,\ i=2,\ 3,\  4$ listed in \cite{ABS}. The AB-KdV systems \eqref{19} with the conditions \eqref{25} can be considered as special reductions of the coupled KdV system derived from two-layer fluid systems \cite{CKdV}.

In addition to the AB-KdV system \eqref{19}, there are other types of AB-KdV systems which can not be directly solved out from the KdV equation. For instance, the following AB-KdV system 
\begin{eqnarray} 
A_t-\frac12A_{xxx}+\frac32B_{xxx}
-3(A-B)A_x+6AB_{x}=0,\quad B= A^{{P}_s{T}_d}.
\label{AKdV1}
\end{eqnarray}
 can be obtained from the reduction of coupled KdV systems, say, the Hirota-Satsuma system \cite{HS}. The Lax pair of the AB-KdV system \eqref{AKdV1} can be directly read out from that of the Hirota-Satsuma system (where $A=u+v,\ B=u-v$) \cite{Fordy,ABS}.

The most general real AB-KdV systems may have the more general form 
\begin{eqnarray}
 K(A,\ B)=0,\ B=\hat{P}_s\hat{T}_dA =A(-x+x_0,\ -t+t_0),\label{KAB}
\end{eqnarray}
with the condition 
$$
 K(u,\ u)=KdV,$$ 
where $KdV$ is defined in Eq. \eqref{kdv}. 

Now a natural question is how to solve the AB-KdV systems listed in this paper. To solve complicated nonlinear systems, the symmetry methods play essential roles. In fact, the powerful Darboux and B\"acklund transformations are essentially special symmetry approaches \cite{DTBT}.  Usually, one uses continuous symmetries to find group invariant solutions of nonlinear systems. 
Here we have derived many AB-systems by using discrete symmetries, the shifted parity and the delayed time reversal. Thus it is interesting that if we can find some nontrivial shifted parity and delayed time reversal invariant solutions for all the AB-KdV systems \eqref{KAB}. 
The group invariant condition 
$
B=\hat{P}_s\hat{T}_dA =A$ 
 implies that to look for the 
$\hat{P}_s\hat{T}_d$ invariant solutions of the AB-KdV systems \eqref{KAB} is equivalent to find the  
$\hat{P}_s\hat{T}_d$ invariant solutions of the usual KdV equation \eqref{kdv}.

For nonlinear systems, soliton excitations are most important exact solutions. For the KdV equation \eqref{kdv}, its well known multiple soliton solutions possesses the form \cite{Hirota} 
 \begin{eqnarray}
u=2\left(\ln F \right)_{xx},\quad 
F=\sum_{\mu}\exp\left(\sum_{j=1}^N\mu_j\xi_j+\sum_{1\leq j<l}^N\mu_j\mu_l \theta_{jl}\right),\label{SolAF1}
\end{eqnarray}
where the summation of $\mu$ should be done for all  permutations of $\mu_i=0,\ 1, \ i=1,\ 2\ \ldots,\ N$ and 
\begin{eqnarray}
\xi_j=k_jx-k_j^3t+\xi_{0j},\quad \exp(\theta_{jl})=\left(\frac{k_j-k_l}{k_j+k_l}\right)^2.  \label{theta1a}
\end{eqnarray}

It is clear that the solution \eqref{SolAF1} is not $\hat{P}_s\hat{T}_d$ invariant. The reason is that the KdV equation is space-time translation invariant and then every soliton of the KdV equation can  be located at anywhere $\xi_{0j}$. However, the AB-KdV is space-time translation symmetry breaking. Thus we have to fix the arbitrary constants $\xi_{0j}$ such that the multiple soliton solution is $\hat{P}_s\hat{T}_d$ invariant but not the space-time translation symmetry invariant. 

After finishing some detailed  calculations, we find that if we rewrite $\xi_j$ as
\begin{eqnarray}
\xi_j=k_j\left(x-\frac{x_0}2\right)-k_j^3\left(t-\frac{t_0}{2}\right)+\eta_{0j}-\frac12\sum_{i=1}^{j-1}\theta_{ij}-\frac12\sum_{i=j+1}^{N}\theta_{ji}
\equiv \eta_{j}-\frac12\sum_{i=1}^{j-1}\theta_{ij}-\frac12\sum_{i=j+1}^{N}\theta_{ji},\label{eta}
\end{eqnarray}
the soliton solution of the KdV equation can be equivalently rewritten as
\begin{eqnarray}
u&=&2\left[\ln \sum_{\nu}K_\nu \cosh\left(\frac12 
\sum_{j=1}^N \nu_j\eta_j 
\right) \right]_{xx},\label{SolA1ch}
\end{eqnarray}
where the summation of $\nu=\{\nu_1,\ \nu_2,\ \ldots,\ \nu_N\}$ should be done for all  permutations of $\nu_i=1,\ -1, \ i=1,\ 2\ \ldots,\ N$, and 
$
 K_\nu=\prod_{i>j}(k_i-\nu_i\nu_jk_j). $

Now, it is straightforward to see that 
\begin{equation}
A=u|_{\eta_{0j}=0}
\end{equation}
solves all the AB-KdV system \eqref{KAB} including Eqs. \eqref{19} and \eqref{AKdV1}. 

Similarly, all the $\hat{P}_s\hat{T}_d$ invariant solutions of the KdV equation are also  solutions of all AB-KdV systems. Here we list two more examples, the Painlev\'e II reduction and the soliton-cnoidal periodic wave interaction solutions. 

The $\hat{P}_s\hat{T}_d$ invariant Painlev\'e II reduction possesses the form,
\begin{equation}
A=\left(t-\frac{t_0}2\right)^{-\frac23}U(\xi)^2+\frac16\frac{2x-x_0}{2t-t_0},\ \quad \xi=\left(x-\frac{x_0}2\right)\left(t-\frac{t_0}2\right)^{-\frac13}, 
\label{su}
\end{equation}
where $U=U(\xi)$ satisfies 
$$
U_{\xi\xi}+U^3+\frac16\xi U+\alpha U^{-3}=0$$
which is equivalent to the Painlev\'e II equation. 

A simple $\hat{P}_s\hat{T}_d$ invariant soliton-cnoidal periodic wave interaction solution \cite{NonL} can be written as 
\begin{eqnarray}
&&A=2w_{xx}\tanh (w) +\frac{w_x^2}6\tanh^2(w)-\frac{w_t}{6w_x}+\frac23\frac{w_{xxx}}{w_x}-\frac12\frac{w_{xx}^2}{w_x^2}-\frac43w_x^2,
\end{eqnarray}
where, $ w=\frac12 k \xi_1 \pm\frac12\mbox{\rm arctanh}\left(m\mbox{\rm sn}(k \xi_2,m)\right),$ $ k=\sqrt{\frac{v_1-v_2}{2(1-m^2)}}$, $\xi_1=\left(x-\frac{x_0}2\right)-v_1\left(t-\frac{t_0}2\right),$ $ 
\xi_2=\left(x-\frac{x_0}2\right)-v_2\left(t-\frac{t_0}2\right) $
with arbitrary constants $v_1,\ v_2$ and $m$.

It should be also emphasized that though
we have obtained many $\hat{f}$ invariant solutions for many models, there are various other solutions 
which are $\hat{f}$ symmetry breaking. Usually, $\hat{f}$ symmetry breaking solutions may not be same for different models. Here we just write down an $\hat{f}$ symmetry breaking soliton solution for the AB-KdV equation \eqref{AKdV1},
\begin{eqnarray}
A=a\tanh \zeta -2k^2\tanh^2\zeta 
+k^2-\frac14\frac{a^2}{k^2},\ \quad 
\zeta=k\left(x-\frac{x_0}{2}\right)-\frac{3a^2-4k^4}{2k}\left(t-\frac{t_0}{2}\right),
\end{eqnarray}
where $k$ and $a$ are arbitrary constants. More about $\hat{f}$ symmetry breaking solutions will be reported in our near future studies. 

\section{Derivation and application of a special AB-KdV system in atmospheric dynamics} 
In Ref. \cite{Jia}, Jia at al. established a multiple vortex interaction model 
\begin{eqnarray}
&&\omega_{i}=\psi_{ixx}+\psi_{iyy}, \\
&&\omega_{it}+[\psi_i,\ \omega_i]+\epsilon\sum_{j\neq i}^N[\psi_i,\ \omega_j]-C\sum_{j=1}^N [\omega_i,\ \omega_j]+\beta \psi_{ix}=0, \label{VVIM}
\end{eqnarray}
where $[\psi_i,\ \omega_j]\equiv \psi_{ix}\omega_{jy}-\psi_{iy}\omega_{jx},\ j\neq i$ denotes the $i$-$j$th stream-vorticity interactions (SVI) and $[\omega_i,\ \omega_j]$ denotes the $i$-$j$th vorticity-vorticity interactions (VVI), $C$ is related to the strength of VVI and the $\beta$ term comes from the coriolis force. In \eqref{VVIM},  the small parameter $\epsilon$ is introduced  by considering the fact that the stream-vorticity interaction
between two faraway (both in space and time) events should be small.

For $N=1$, the model \eqref{VVIM} is just the well known (2+1)-dimensional rotating fluid model (Euler equation with rotating effect), such as the atmospheric and oceanic systems, which can be used to describe the nonlinear inviscid dissipative and equivalent barotropic vorticity equation (NIDEBE) in a $\beta$-plane channel \cite{beta}. The model \eqref{VVIM} is also derived from NIDEBE by neglecting higher order smaller interactions among different vortices. 

In fact, the model can also be used to describe $N$-event problems. For the $N=2$ case, Eq. \eqref{VVIM} becomes
\begin{eqnarray}
&&\omega_{i}=\psi_{ixx}+\psi_{iyy},\quad i=1,\ 2,
\label{VVIM0}\\
&&\omega_{1t}+[\psi_1,\ \omega_1]+\epsilon[\psi_1,\ \omega_2]-C [\omega_1,\ \omega_2]+\beta \psi_{1x}=0, \label{VVIM1}\\
&&\omega_{2t}+[\psi_2,\ \omega_2]+\epsilon[\psi_2,\ \omega_1]-C [\omega_2,\ \omega_1]+\beta \psi_{2x}=0. \label{VVIM2}
\end{eqnarray}
It is clear that the model \eqref{VVIM0}--\eqref{VVIM2} allows an AB reduction, the AB equivalent barotropic vorticity equation (AB-EBVE)
\begin{eqnarray}
&&\psi_2=\hat{f}\psi_1=\hat{P}_s^x\hat{T}_d\psi_1=\psi_1(-x+x_0,\ y,\ -t+t_0),\nonumber\\
&&\omega_2=\hat{f}\omega_1=\hat{P}_s^x\hat{T}_d\omega_1=\omega_1(-x+x_0,\ y,\ -t+t_0).\label{ABVVI}
\end{eqnarray}
To get some approximate analytic solution of the AB-EBVE, we utilize the multiple scale method (MSM) to derive an AB-KdV system from Eqs.
\eqref{VVIM0}--\eqref{VVIM2} with conditions \eqref{ABVVI}. 

As in the standard MSM, by introducing some slow variables,
\begin{eqnarray}
\xi=\epsilon^{1/2}(x-ct), \ \tau=\epsilon^{3/2} t, \label{xt}
\end{eqnarray}
the stream function $\psi_1$ can be expanded as 
\begin{eqnarray}
\psi_1=u_0(y)-cy+A_1(\xi,\ y,\ \tau)\epsilon + A_2(\xi,\ y,\ \tau)\epsilon^2 +O(\epsilon^3), \label{psi1}
\end{eqnarray}
while the model parameters, $C$ and $\beta$, can also be expanded in some series of $\epsilon$. In this paper, we just take 
\begin{eqnarray}
C=\delta \epsilon,\quad \beta=\mu \epsilon^2 \label{bc}
\end{eqnarray}
which means for two far away events, the vorticity-vorticity interaction is in the same order as for the stream-vorticity interaction while the effect of the coriolis force is smaller in the next order. 

Substituting the expansion
\eqref{psi1} with Eqs. \eqref{ABVVI} and \eqref{bc} into the original model
\eqref{VVIM1} and \eqref{VVIM2}, we have
\begin{eqnarray}
&&\left(u_{0yyy}A_{1}-u_{0y}A_{1 yy}\right)_{\xi}\epsilon^{3/2}+\left[\big(u_{0yyy}A_{2}-u_{0y}(A_{2 yy}+A_{1\xi\xi})\big)_{\xi}+(\delta u_{0yy}-A_1)_yA_{1\xi yy}
\right.
\nonumber\\
&& \left.
-[(\delta u_{0yy}+u_0)_y-c] B_{1\xi yy}+(u_{0yyy}+A_{1yyy}+\mu)A_{1\xi}+A_{1\tau yy}\right]\epsilon^{5/2}=O(\epsilon^{7/2}),
\label{K0}\\
&&\left(u_{0yyy}B_{1}-u_{0y}B_{1 yy}\right)_{\xi}\epsilon^{3/2}+\left[\big(u_{0yyy}B_{2}-u_{0y}(B_{2 yy}+B_{1\xi\xi})\big)_{\xi}+(\delta u_{0yy}-B_1)_yB_{1\xi yy}
\right.
\nonumber\\
&& \left.
-[(\delta u_{0yy}+u_0)_y-c] A_{1\xi yy}+(u_{0yyy}+B_{1yyy}+\mu)B_{1\xi}+B_{1\tau yy}\right]\epsilon^{5/2}=O(\epsilon^{7/2}),
\label{K0B}
\end{eqnarray}
where 
\begin{eqnarray}
&&B_i\equiv \hat{P}^{\xi}_s\hat{T}_d^{\tau}A_i =A_i(-\xi+\xi_0,\ y,\ -\tau+\tau_0),\ i=1,\ 2,\ (\xi_0=\epsilon^{1/2}(x_0-ct_0),\ \tau_0=\epsilon^{3/2}t_0). \label{Bi}
\end{eqnarray}
In the subsequent steps we will only treat Eq. \eqref{K0} because Eq. \eqref{K0B} is only an AB dual of Eq. \eqref{K0}.

Eliminating the leading term (the term with $\epsilon^{3/2}$) of Eq.  \eqref{K0}, 
we have 
\begin{eqnarray}
\left(u_{0yyy}A_{1}-u_{0y}A_{1 yy}\right)_{\xi}=0. \label{D1}
\end{eqnarray}
It is clear that Eq.  \eqref{D1}
can be solved via variable separation 
\begin{eqnarray}
A_1=f(y)A(\xi,\ \tau)\equiv fA,\quad \left(\mbox{and}\quad B_1=fB=f\hat{P}^{\xi}_s\hat{T}_d^{\tau}A\right). \label{VS}
\end{eqnarray}
Substituting Eq. \eqref{VS} into Eq. \eqref{D1} yields 
\begin{eqnarray}
u_{0yyy}f-u_{0y}f_{yy}=0
 \label{Df}
\end{eqnarray}
with a special solution 
\begin{eqnarray}
&&f=c_0u_{0y} \label{rU0}
\end{eqnarray}
where $c_0$
is an arbitrary constant.

Vanishing the coefficient of $\epsilon^{5/2}$ in Eq. \eqref{K0} and using the relations \eqref{VS} and \eqref{rU0}, we have
\begin{eqnarray}
&&\big(u_{0yyy}A_{2}-u_{0y}(A_{2 yy}+A_{1\xi\xi})\big)_{\xi}+(\delta u_{0yy}-A_1)_yA_{1\xi yy}
\nonumber\\
&& 
-[(\delta u_{0yy}+u_0)_y-c] B_{1\xi yy}+(u_{0yyy}+A_{1yyy}+\mu)A_{1\xi}+A_{1\tau yy}=0,
\label{K1}
\end{eqnarray}
To solve $A_2$ from \eqref{K1}, we can take the following variable separation form 
 \begin{eqnarray}
A_2=u_{0y}\left(u_1A+u_2A^2+u_3A_{\xi\xi}+u_4B_{\xi\xi}+u_5B^2+u_6B+u_7AB\right)
\label{A2}
\end{eqnarray}
with $u_i,\ i=1,\ 2,\ \ldots,\ 7$ being arbitrary functions of $y$. 

Substituting Eq. \eqref{A2} into Eq. \eqref{K1} leads to
\begin{eqnarray}
&&c_0^2f_{yy}A_{\tau}-\left[\big(f^2u_{7y}\big)_yA+2\big(f^2u_{5y}\big)_yB+\big(f^2u_{6y}\big)_y+c_0f_{yy}\big(f+\delta f_{yy}-cc_0\big)
\right]B_{\xi}\nonumber\\
&&
-\left[\big(2f^2u_{2y}+c_0^2f_y^2-c_0^2ff_{yy}\big)_yA+\big(f^2u_{7y}\big)_yB+\big(f^2u_{1y}\big)_y-c_0f(c_0\mu+f_{yy})-c_0\delta
f_{yy}^2\right] A_{\xi}\nonumber\\
&&-\left[\big(f^2u_{3y}\big)_y+c_0f^2\right]A_{\xi\xi\xi}-\big(f^2u_{4y}\big)_yB_{\xi\xi\xi}
=0.\nonumber\\
\label{KdV0}
\end{eqnarray}
Finally, as in the usual multiple scale method, taking an average for the fast variable $y$ over the whole $\beta$-channel (from $y_1$ to $y_2$), i.e., applying 
$$\frac{1}{y_2-y_1}\int_{y_1}^{y_2}\mbox{d}y$$ 
on Eq. \eqref{KdV0}, we obtain the following AB-KdV system
\begin{eqnarray}
&&A_{\tau}+\alpha_1 A_{\xi}+\alpha_2 AA_\xi+\alpha_3 A_{\xi\xi\xi}+\alpha_4 B_{\xi\xi\xi}+\alpha_5 BB_{\xi}+\alpha_6 B_\xi
+\alpha_7 (BA)_{\xi}
=0,
\label{KdVa}\\
&&
 B=\hat{P}_s^{\xi}\hat{T}_d^\tau A,
\label{KdVB}
\end{eqnarray}
where 
\begin{eqnarray}
&&\alpha_0=f_{y}(y_2)-f_y(y_1)\equiv \left.f_y\right|_{y_1}^{y_2},\
\alpha_1=-\frac{1}{\alpha_0 c_0^2}\left[\left.\big(f^2u_{1y}\big)\right|_{y_1}^{y_2}
-c_0\int_{y_1}^{y_2}\left(c_0\mu f+ff_{yy}+\delta f_{yy}^2\right)\mbox{d}y\right],
\nonumber\\
&&
\alpha_2=\frac{1}{\alpha_0 c_0^2}\left.\big[c_0^2(ff_{yy}-f_y^2)-2f^2u_{2y}\big]\right|_{y_1}^{y_2},\
\alpha_3=-\frac{1}{\alpha_0 c_0^2}\left[\left.\big(f^2u_{3y}\big)\right|_{y_1}^{y_2}+c_0\int_{y_1}^{y_2}f^2\mbox{d}y\right],
\nonumber\\
&&
 \alpha_4=-\frac{1}{\alpha_0 c_0^2}\left.\big(f^2u_{4y}\big)
 \right|_{y_1}^{y_2},\
 \alpha_5=-2\frac{1}{\alpha_0 c_0^2}\left.\big(f^2u_{5y}\big)\right|_{y_1}^{y_2},\ 
\nonumber\\
&&\alpha_6=-\frac{1}{\alpha_0 c_0^2}
\left[\left.\big(f^2u_{6y}-cc_0^2f_y\big)\right|_{y_1}^{y_2}
+c_0\int_{y_1}^{y_2}f_{yy}\left(f+\delta f_{yy} \right)\mbox{d}y\right],\
\alpha_7=-\frac{1}{\alpha_0 c_0^2}\left.\big(f^2u_{7y}\big)\right|_{y_1}^{y_2}.
\end{eqnarray}

Thus, the AB-EBVE \eqref{VVIM0}-\eqref{VVIM2} with the condition \eqref{ABVVI} possesses the approximate solution 
\begin{eqnarray}
\psi_1&=&u_0-cy+u_{0y} \epsilon \left[c_0A
+\epsilon \left(u_1A+u_2A^2+u_3A_{\xi\xi}+u_4B_{\xi\xi}+u_5B^2+u_6B+u_7AB\right)\right],\label{p1}\\
\psi_2&=&\hat{P}_s^x\hat{T}_d\psi_1
=\hat{P}_s^{\xi}\hat{T}_d^{\tau}\psi_1,\label{p2}
\end{eqnarray}
where $u_i,\ i=0,\ 1,\ 2,\ \ldots, 7 $ are  arbitrary functions of $y$, and $A$ and $B$ are solutions of the AB-KdV system \eqref{KdVa}.  

If $B=A$, the AB-KdV equation \eqref{KdVa} will return to the standard KdV equation \eqref{kdv} after some suitable scaling and Galileo transformations. Then, the AB-KdV equation \eqref{KdVa} possesses $\hat{P}_s^{\xi}\hat{T}_s^{\tau}$-invariant $N$-soliton solutions, Painlev\'e II reductions and interaction solutions between cnoidal periodic wave and soliton as  mentioned in the last section. 

In fact, there exist also some $\hat{P}_s^{\xi}\hat{T}_s^{\tau}$ symmetry breaking solutions. For instance, it is straightforward to verify that the AB-KdV system \eqref{KdVa} with the condition
$$\alpha_7=\frac12\frac{(5\alpha_3+7\alpha_4)\alpha_2-\alpha_5(5\alpha_4+7\alpha_3)}{\alpha_3-\alpha_4} $$ 
possesses the following periodic wave solution 
\begin{eqnarray}
&&A=a_0+a_1\mbox{sn}\left(\Xi,\ m\right)
+\frac{2m^2k^2(\alpha_3-\alpha_4)}
{\alpha_5-\alpha_2}\mbox{sn}^2\left(\Xi,\ m\right),\label{sn}\\
&&\Xi\equiv k\left(\xi-\frac12\xi_0\right)+
\omega\left(\tau-\frac12\tau_0\right) 
\nonumber\\
&&\omega=\frac{2k^3(1+m^2)(\alpha_3^2-\alpha_4^2)}{5\alpha_3+7\alpha_4}+\frac{(7\alpha_3+5\alpha_4)\alpha_6 k}{7\alpha_4+5\alpha_3}-\alpha_1 k+\frac{(\alpha_2-\alpha_5)[(\alpha_3+2\alpha_4)\alpha_2-(2\alpha_3+\alpha_4)\alpha_5]a_1^2}{(5\alpha_3+7\alpha_4)(\alpha_3-\alpha_4)m^2k},\nonumber\\
&& a_0=\frac{k^2m^2(\alpha_3-\alpha_4)\big[k^2(m^2+1)(3\alpha_3+5\alpha_5)-2\alpha_6\big]}{(5\alpha_3+7\alpha_4)(\alpha_2-\alpha_5)m^2k^2}-\frac{a_1^2
\big[\alpha_2(\alpha_3+2\alpha_4)-\alpha_5(\alpha_4+2\alpha_3)\big]}{(5\alpha_3+7\alpha_4)(\alpha_3-\alpha_4)m^2k^2}\nonumber
\end{eqnarray}
with the arbitrary constants $a_1$, $k$, $m$ and the Jacobi elliptic sine function, $\mbox{sn} (\Xi,\ m)$. It is obvious that the solution \eqref{sn} is $\hat{P}_s^{\xi}\hat{T}_s^{\tau}$ invariant only for $a_1=0$. 
The periodic wave solution \eqref{sn} will be reduced to a $\hat{P}_s^{\xi}\hat{T}_s^{\tau}$ symmetry breaking soliton solution for $m^2\rightarrow 1$ and $a_1\neq 1$. 

The atmospheric and oceanic phenomena are rich and multifarious. The richness and multifariousness of our approximate solution comes from the introduction of arbitrary functions $u_0,\ u_1,\ \ldots,\ u_7$. By selecting the arbitrary functions appropriately, we can explain various two-place correlated events in atmospheric and oceanic dynamics. For example, Fig. 1a displays the theoretic result of A-event (atmospheric blocking) described by $\psi_1$ of \eqref{p1} happened near 40 degree east longitude while the Fig. 1b is the corresponding real weather chart of geopotential height at 500 hPa on 19th November, 2007, from the National Centers for Environmental Prediction/the National Center for Atmospheric Research (NCEP/NCAR) reanalysis data\cite{HF}. Fig. 2a shows the theoretic density and contour plot of the B-event described by $\psi_2$ of \eqref{p2} appeared near 130 degree west longitude while the Fig. 2b is the corresponding plot of the real weather chart happened after 65 days later after 
the A-event. The function and parameter selections related Figs. 1a and 2a are as follows,
\begin{eqnarray}
&& u_0=C_0+\frac{f_0}{c_0k_1}\tanh(k_1y-y_0),\ (f=f_0\mbox{sech}^2(k_1y-y_0),\\
&& u_i=C_i \sin(k_2(y-y_3)),\ i=1,\ 2,\ \ldots,\ 7
\end{eqnarray}
with 
\begin{eqnarray}
&&C_0=m=\mu=k=1,\ C_1=C_3=0.1,\ C_2=-10.43389,\ C_4=C_5=\xi_0=\tau_0=0, \nonumber\\
&&C_6=2480,\ C_7=-14.7,\ a_1=c=0.01,\ c_0=10,\ f_0=-0.1,\ y_0=3,\ y_1=50,\nonumber\\
&& y_2=70,\ y_3=40,\ k_1=k_2=\frac1{15},\ \delta_0=\frac16,\
 \epsilon=\frac1{140}, \label{cc}
\end{eqnarray}
$t=t_A=12 \ (days) $ for A-event and $t=t_B=12+65=77\ (days) $ for B-event. 

The long time real atmospheric blockings evolving from November 2007 to January 2008 including the A and B events mentioned above may be responsible for the 
heavy snow disaster of South China in the winter of 2007/2008. The blocking occurred at the north Europe in November 2007, which located at the upstream of China, may favor to induce cold air outbreaks to China from Barents Sea and Kara Sea of the Arctic by northwest wind in front of the blocking high ridge (Fig. 1b).  After the cold surge which caused the heavy snow disaster in the south 
of China in the winter 2007/2008, energy dispersed downstream and new blocking was developed in eastern Pacific-north American region (Fig. 2b).\\  
\\
\centerline{\includegraphics[width=10cm]{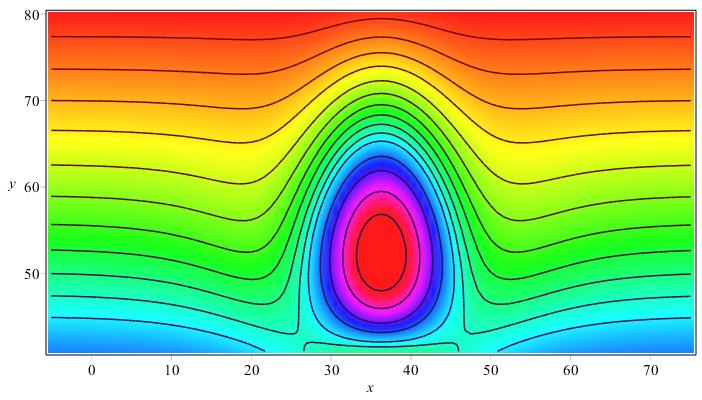}}

\centerline{(a)}

\centerline{\includegraphics[width=10cm]{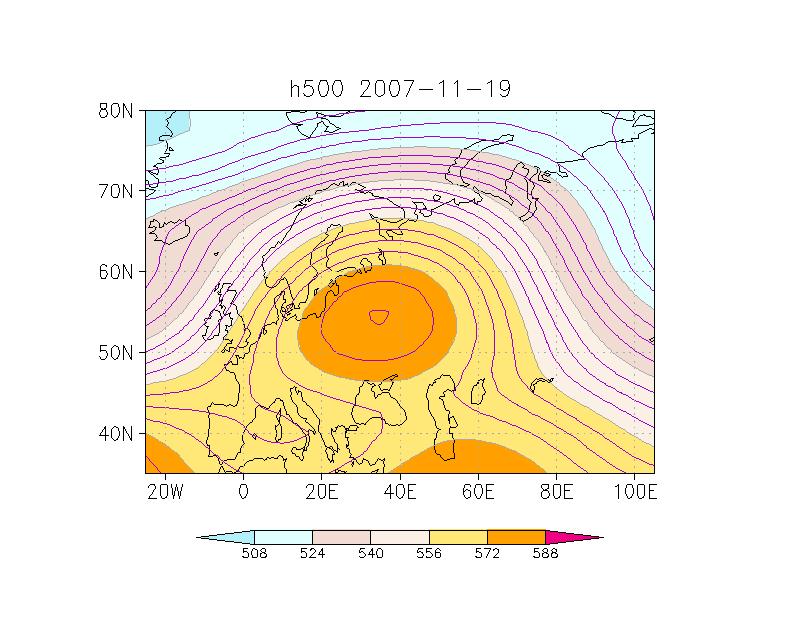}}
 
\centerline{(b)}

\centerline{\footnotesize\ Figure 1 (a) the theoretic result of A-event (atmospheric blocking)
described by $\psi_1$ of (51) and (b)}  
\centerline{\footnotesize  the corresponding real
weather chart of geopotential height at 500 hPa on 19th November, 2007. }

\centerline{\includegraphics[width=10cm]{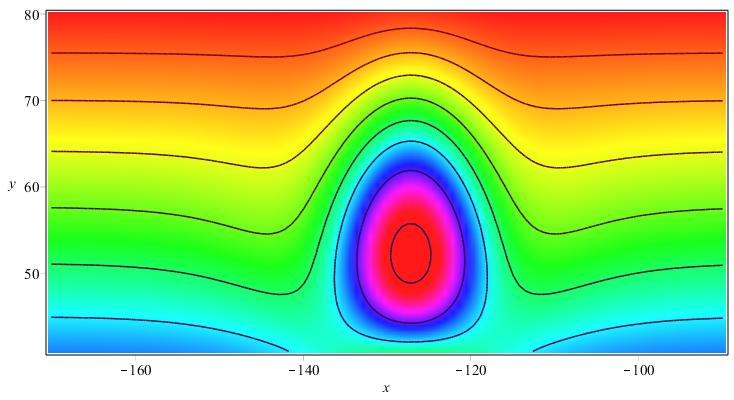}}

\centerline{(a)}

\centerline{\includegraphics[width=10cm]{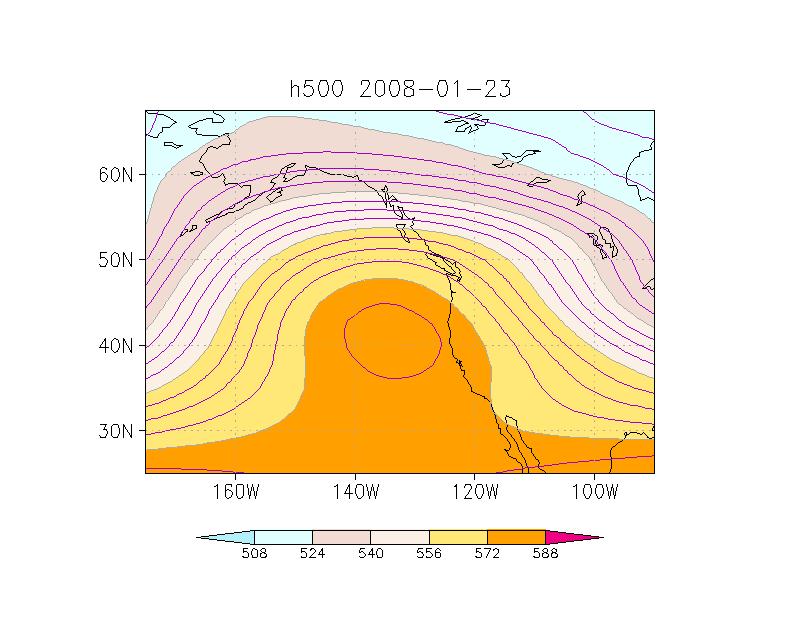}}
 
\centerline{(b)}

\centerline{\footnotesize Figure 2 (a) the theoretic density and contour plot of the B-event described by $\psi_2$ of (52)
 }  
\centerline{\footnotesize  and (b) the corresponding plot of the real weather chart
happened at 23rd January 2008.}

\section{Summary and discussions}   

In summary, 
a new physical fields, AB-Physics or even AB-Science, is opened. This is a first attempt to study two-place physics and multi-place physics, which may be widely and deeply affect all other scientific fields. The reason to introduce Alice-Bob physics is because there exist various two-place (and multi-place) correlated physical events in almost all natural scientific fields. For instance, all the teleconnection phenomena in  atmospheric and oceanic science 
belong to the 
 category of the AB-physics/science.

A simple method is established to find huge amount of new physically meaningful nonlinear models, Alice-Bob systems. The method can be applied to any known physics to find novel phenomena though here we take only the celebrating KdV model as an illustration to find infinitely many Alice-Bob KdV systems. 
A special type of AB-KdV systems, the functional KdV equations,  with an arbitrary $\hat{P}_s\hat{T}_d$ invariant functional $G(A,\ B)$, can be derived directly from the original KdV system. Some other types of AB-KdV equations can be directly obtained from coupled KdV systems by using correlated relations between two components. The method can be applied to any original principle physical models such as the Maxwell equations, Navier-Stokes equations, 
Schr\"odinger equations, 
Einstein field equation 
and their integrable and nonintegrable derivative systems. In the preprint paper \cite{ABS}, various other types of 
integrable AB systems have been listed by directly applying suitable correlated relations on coupled two component models.    

A special AB-KdV system is derived from the multiple vorticity interaction model which is related to a standard atmospheric and oceanic dynamic system, the nonlinear inviscid dissipative and barotropic vorticity equation in a $\beta$-plane channel. 
The $\hat{P}_s\hat{T}_d$ symmetry breaking soliton solution of the derived AB-KdV system is used to qualitatively describe the two real events, the atmospheric blocking happened in November 2007 and January 2008 respectively while the atmospheric blockings are responsible for the heavy snow disaster in Southern China in the winter 2007/2008.

Some types of physically important new symmetries such as shifted parity and delayed time reversal, are found. It is well known that parity and time reversal are two very important symmetries in physics. Now these two symmetries are extended to more general symmetries. These new symmetries exist in various physical fields which include not only the fields where the KdV equation is valid but also those where the Maxwell equations, (linear and nonlinear) Schr\"odinger equations, Navier-Stokes equation and Einstein equation are valid. In addition to the shifted parity and delayed time reversal symmetries, we have found more newly general symmetries given in theorem 1. These symmetries will be very useful in two place physics and should be studied further.

Some types of group (shifted parity and delayed time reversal symmetry group) invariant solutions including N-soliton solutions (arbitrary N), Painlev\'e II reductions and soliton cnoidal wave interactions are obtained for all AB-KdV systems. The physical meaning of the group invariant solutions is that the event happened at $\{x,\ t\}$ will happen also at $\{x',\ t'\}$.

One special shifted parity ($\hat{P}_s$) and delayed time reversal ($\hat{T}_d$) symmetry breaking soliton solution is also given for special AB-KdV systems \eqref{19} and \eqref{KdVa}. The physical meaning of the group symmetry breaking solution is that for Alice-Bob systems there are real physical phenomena where the event B at $\{x',\ t'\}$ is different from the event A at $\{x,\ t\}$.
 
Infinitely many nonlinear excitations are found to be solutions of infinitely many models. This fact indicates and emphasizes that to conclude a theory 
	should be very careful even if you have observed various (may be infinitely many) facts are valid for a theory without any parameters. For instance, even if you have observed N-soliton solutions for an arbitrary N in a real system, there are still infinitely many candidate theories to describe this system. 
	
It is indicated that additional nonlinearities can be introduced by shifted parity and delayed time reversal correlated constrained conditions. That means shifted parity and delayed time reversal correlation conditions are nonlinear constraints. This implies that we can introduce useful nonlinearities from useful nonlocal nonlinear symmetry invariant constraints. In the reference \cite{Nothing}, one of us, Lou, prove that various physically important nonlinear systems (such as KdV, KP, NLS, sine-Gordon etc.) can be derived from linear systems by means of this types of nonlinear and nonlocal constraints . 
	
It is also indicated that there are infinitely many nonlocal physical functional models which have not yet been studied by present physics and lack of effective methods to solve nonlocal and nonlinear functional physical models. This fact arouses many challenges to the present physical society. 
	
Though the multiple soliton solutions for many integrable systems have been obtained in some types of formally different complicated expressions, in this paper, a much more elegant form for the KdV type systems is proposed. The similar elegant forms for many other types of soliton systems can also be obtained\cite{ABS} by using the same approach.
 
 In addition to the beauty of the new soliton expressions, our forms display two new physical phenomena. The first one is that the resonant soliton solution is analytical when the wave numbers $k_i$, and $k_j$ are very closed to each other, however, it is not analytical for the traditional multiple soliton expressions when resonance happens. The second one is that if we take the positions of the solitons are wave numbers dependent, then we may obtain various types of quite different resonant solutions which need to be further explored in  future.

 \bf Acknowledgement. \rm
The authors is in debt to Profs. J. F. He, Z. N. Zhu, Y. Chen, J. Lin, D. J. Zhang, B. F. Feng, X. B. Hu, Q. P. Liu, Y. Q. Li and X. Y. Tang for their helpful discussions. The work was sponsored by the Global Change Research
Program of China (No.2015CB953904), the National Natural Science Foundations of China (Nos. 11435005), Shanghai Knowledge Service Platform for Trustworthy Internet of Things (No. ZF1213) and K. C. Wong Magna Fund in Ningbo University.


\end{document}